\documentclass[3p]{elsarticle}

\biboptions{sort&compress}

\usepackage[utf8]{inputenc} 
\usepackage[T1]{fontenc}    
\usepackage{hyperref}       
\usepackage{url}            
\usepackage{amsmath, amssymb, amsfonts, bm, stmaryrd, soul}       
\usepackage{graphicx, xcolor, caption, subcaption}       

\begin{document}

\begin{frontmatter}
\title{Closing a catenary loop: \\the lariat chain, the string shooter, and the heavy \emph{elastica}}

\author[1]{A. R. Dehadrai}
\ead{d\_abhinav@blr.amrita.edu}
\affiliation[1]{organization={Amrita School of Artificial Intelligence, Amrita Vishwa Vidyapeetham},
             city={Bengaluru},
             country={India}}

\author[2]{J. A. Hanna} 
\ead{jhanna@unr.edu}
\affiliation[2]{organization={Department of Mechanical Engineering, University of Nevada},
             city={Reno},
             postcode={89557-0312},
             state={NV},
             country={U.S.A.}}

\date{\today} 

\begin{abstract}
We review, critique, and extend results related to the problem of closed loop shape equilibria of a string shooter, a type of catenary consisting of steady, axially moving configurations of an inertial, inextensible, perfectly flexible string in the presence of gravity and drag forces. 
We relate to similar problems, including the lariat (no gravity), chain fountain (not closed), and heavy \emph{elastica} (bending stiffness). We focus on the difficulty inherent to continuing a catenary through a vertical orientation, necessary to close a loop, which difficulty changes in nature as the system undergoes bifurcations with increasing drag. We construct solutions by implementing available analytical results, and numerically generate additional solutions with added bending stiffness.  We briefly discuss global balances of linear, angular, and pseudo- momentum for this system. 
\end{abstract}

\end{frontmatter}

\section{Introduction}

This extended comment, including a few new results, was stimulated by recent academic attention directed at a ``cute'' but challenging and historied problem in string mechanics, arising from its inclusion as part of problem 13 of the 11th International Physicists' Tournament 
\cite{IPT19}, in turn inspired by a demonstration by Yeany \cite{YeanyYouTube}. The problem is to determine the shape equilibria of a ``string shooter'', a closed loop of axially moving string with a single support, subject to gravity and drag forces, to be described in detail below. 
As will be discussed at length, several important aspects of this and related problems, already documented throughout the literature, have been overlooked, 
and some recently 
published numerical solutions to the problem are not physically realizable.  
These conclusions follow from inspection of prior analytical results \cite{Gregory49, SvetlitskiiMiroshnik73, Miroshnik01, ChakrabartiHanna16}, however, errors in the 
presentation of the results in \cite{ChakrabartiHanna16} must also be corrected in order to construct solutions. 

We will provide a perspective on catenary problems that reviews both very old and somewhat new results, emphasizes a fundamental barrier to constructing solutions in the form of a closed loop with insufficient drag forces, and calls attention to the curious forms of singularities arising in this class of problems.  
Along the way, we must navigate past several distractions 
 to reach the crux of the physics. 
    In particular, from our perspective, the importance of drag has been misunderstood, and perhaps over-emphasized. Although 
    sufficient drag is necessary to construct loop solutions, its role is to change tension and curvature profiles, at times through bifurcations. 
 On this point, classical results for axially moving bodies in the absence of drag, such as catenaries or free ``lariats'', are illustrative. 

When drag forces are moderate to large, we indicate how analytical solutions can be used in a semi-analytical framework to construct closed loops. 
When drag forces are lower, such solutions are unphysical, and we offer a path towards numerical resolution of the problem using additional bending stiffness. This includes the problem of a static heavy \emph{elastica} or, equivalently, a stiff catenary. 
Although bending stiffness, which is clearly present in some experiments, 
 is a likely source of regularization, the nature of this regularization takes an atypical form. Instead of a high curvature boundary layer, derivatives of curvature contribute a bending force that allows the body to pass through an orientation aligned with gravity.  We are unaware of any previous examination of this type of regularization.     
The reason we say ``path towards'' resolution is that the resulting problem of a very floppy beam is very stiff numerically, 
making the likely parameter range of experiments difficult to access.  
We show that a small amount of bending has a large qualitative effect on the solutions. 
We do not attempt to match experiment to determine whether bending is indeed the regularizing agent in the laboratory. 

After a verbal introduction to the problem, its history, and some conceptual difficulties and misunderstandings, we proceed to state the problem mathematically, and present and discuss characteristic features of solutions in different physical regimes. 
Practical details are relegated to \ref{practical}.

\section{Overview of the problem and its solutions}\label{problemsummary}

\begin{figure}[h!]
     \centering
     \includegraphics[width=0.45\textwidth]{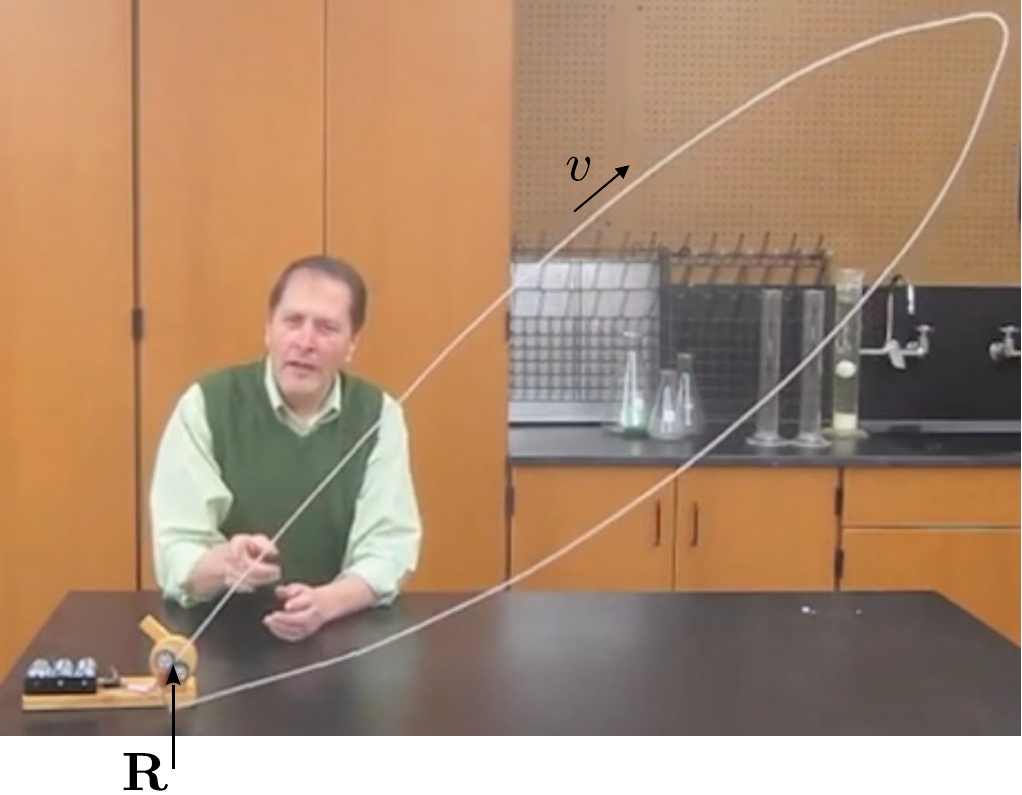}
     \caption{A string shooter in operation, and its inventor \cite{StringLauncher}.  The string moves axially clockwise with speed $v$. The launch mechanism provides a force ${\bf{R}}$. 
     (Annotations added to a still from \cite{YeanyYouTube}, flipped horizontally for consistency with other figures in the present paper.)}
        \label{yeany}
\end{figure}

Briefly, the problem at hand considers observed steady or nearly-steady configurations of an axially moving loop of string or chain subject to gravity and drag forces.  Most of the body is suspended in air, but at one location, one or more motorized wheels drive the axial motion, as well as provide a necessary force to support the structure. 
An example of such a flowing structure is shown in Figure \ref{yeany}. 
For simplicity, we may presume perfect flexibility, inextensibility, and uniform density of a continuous body, coplanar with gravity.
What, if any, are the equilibrium loop shapes and corresponding tensions, and how do they depend on drag forces?  Here by ``loop'' we mean a closed curve that might have kinks (jumps in angle), particularly at the support point. 
The tangents of such a shape must necessarily either have two or more kinks, or subtend an angle greater than $\pi$, including a vertical point (where the normal is horizontal). 
However, jumps in internal force and moment are only admissible at the support point, where external supplies are available. 
We anticipate that the supply of force will be vertical (opposing gravity), and expect that the supply of moment should be zero for a perfectly flexible string. 

The analytical results of \cite{Gregory49, SvetlitskiiMiroshnik73, Miroshnik01, ChakrabartiHanna16} can be used to generate solutions for sufficiently high drag forces. 
For lower drag forces, \emph{there are no solutions} to the problem as posed above. 
  
 To understand why, it is helpful to review 
the classical case of a static catenary, in which tension balances gravity.
The string's contribution to normal force balance is the product of tension and curvature.  There are two consequences of this: any nonvertical element of string cannot be straight or tensionless, and any vertical element of string must either be straight or tensionless or both, otherwise there will be an unbalanced force. 
With the exception of nongeneric vertical straight line solutions, a hanging catenary approaches the vertical asymptotically, with tension and arc length tending to infinity, and curvature tending to zero.  The same is true for a catenary arch, with the tension negative.  
Centripetal acceleration associated with axial motion will shift the tension by some positive constant, which can affect stability, but does not change the shape. 
  
The equations of an axially moving catenary, including inertial and sufficiently weak drag forces, also generically admit only curves with everywhere nonzero curvature and nonvertical orientation, approaching straightness and verticality only asymptotically at infinite length \cite{ChakrabartiHanna16}. 
One might thus seek to patch 
 together two separate catenary solutions--- an upper and a lower branch--- near vertical orientation to create a closed loop.  
Finite-length shapes must necessarily be nonsmooth, with jumps in angle at the patching points that can be made arbitrarily small.  
While this can easily provide visually appealing matches with experimental shapes, indicating that catenary physics captures the dominant balances over much of the domain, such results are not physically admissible. 
The error in this approach is that the tension must also suffer a jump, requiring another external force that is not provided by the physical conditions of the problem.  This jump in tension blows up as the jump in angle shrinks with increasing length of the curves.

 However, stronger drag forces allow for physical patching, through two different scenarios, depending on the magnitude of the drag. As noted in \cite{SvetlitskiiMiroshnik73, Miroshnik01, ChakrabartiHanna16}, although misinterpreted in \cite{ChakrabartiHanna16}, there are three regimes, separated by two bifurcations: 
\begin{itemize}
	\item At low drag to weight ratios, there is a jump in tension that blows up as the curve approaches vertical orientation.  Patching curves at a kink would require an external force at a second, unsupported point, so these solutions are not admissible.  
	\item At moderate drag to weight ratios, it becomes possible, on one side, to patch at finite length at vertical orientation without a kink.  There the inertially shifted tension vanishes, as does the curvature. 
	\item At high drag to weight ratios, on one side, the curvature blows up at finite length at vertical orientation.  The shifted tension still vanishes there, and the curves are rectifiable and have continuous tangents.  
 It is still possible to patch solutions, valid for perfectly flexible strings.  
\end{itemize}

At low drag, a catenary requires two support forces. 
Without a second support force, closed curves are impossible, and such cases have been mostly ignored. Yet this problem smoothly connects to that of a static hanging loop, which is easy to physically realize. This clearly indicates a need for additional physics in the model, likely in the form of bending elasticity. 

The results of patching will be seen in greater detail below.  
The symmetry or asymmetry of the shapes depends on the interrelated launch and return angles of the shot string, but it is a curious property of the catenary equations that the lengths of the upper and lower curve branches will be equal, even for asymmetric solutions \cite{Gregory49}.  A given launch angle and string length will determine a particular spatial location for the patch point, and a particular return angle.

\section{A bit of history}\label{history}

The origin of this problem is unknown to us, but variants appear in the scientific literature as early as the 19th century.  
In many of the early investigations, drag forces are a secondary or neglected consideration.

A notable report is that of the meteorologist Aitken \cite{Aitken1878}, who presents a discussion and numerous drawings of observations of a driven heavy chain either hanging over one wheel (lariat chain) or passing between two wheels (string shooter).
A platform underneath is employed to interfere with the motion, this interruption at times enabling the loop to achieve greater heights by effectively transforming it into a (chain) ``fountain''. 
This is a low drag to weight ratio system. 
Aitken understood that acceleration and tension are the dominant forces, and that these could be perfectly balanced \emph{for any shape} in the absence of gravity. 
Even in the presence of gravity, shapes evolve slowly, with the time scale for relaxation greatly exceeding that for a material element to traverse features of interest through axial motion. 
These slow shape dynamics are the primary feature of Tuck's ``Lariat Chain'' kinetic sculpture developed a century later \cite{LariatChain}, many of which serve as science museum installations.\footnote{The ``lariat chain'' \cite{LariatChain} is an earthly thing that lives in the presence of gravity, but serves well to imperfectly illustrate the behavior of the abstract concept of a ``lariat'' \cite{HealeyPapadopoulos90}, an inertial string loop in the absence of gravity.} 
Aitken also discusses other multi-wheel arrangements, and the phenomenon whereby a freed loop will ``run'' along the platform for some distance. 

Aitken's analysis of his experiments benefited from prior knowledge of relevant mechanics. In his own modest assessment, ``The experiments do not contain much that is new, many of the problems having previously been mathematically wrought out by Sir William Thomson and others'' \cite{Aitken1878}.
The dominant normal force balance between centripetal acceleration and tension was well understood in the 1850s by Routh \cite{Routh55}, who may have picked this up from the Mathematical Tripos. 
Both of these contributions depend linearly on curvature, which can be divided out of the normal force balance, such that no information about shape is retained and any configuration of the axially moving body is a marginally stable equilibrium.
One can alternately recognize that one of two arbitrarily shaped nonlinear wave solutions of an axially moving string is ``frozen'' in the lab frame. 
The only effect of axial motion is to positively shift the value of internal tension.
This is generally true for static shape equilibria in which no drag or Coriolis forces are present. 
In particular, catenaries with axial flow retain their shape, with tension shifted, as noted by Routh in his treatise \cite{Routh55},  
and in contemporary analyses of telegraph-laying by Airy 
 \cite{Airy1858} and Thomson 
 \cite{Thomson1857both}; for a review of the literature on paying out and towing cables with additional drag forces, see \cite{ChakrabartiHanna16} and the brief mention of normal drag in the paragraph below. 
For more on shape-agnosticism, tension shifts, and other interesting discussions, we refer the reader to Routh \cite[Chapter XIII, Articles 594-598]{Routh55}, 
 Gray \cite[Chapter XXII, Sections 11-13]{Gray59}, 
and a short note by Healey and Papadopoulos that extends this  
``classical, but virtually forgotten, result'' \cite{HealeyPapadopoulos90}. 
A striking consequence of the shift in tension due to axial flow is the stabilization of structures lacking bending stiffness that would otherwise collapse if under compression, in particular the catenary arch.  
More than a century after Aitken's demonstrations, this phenomenon was rediscovered, explained, and experimentally demonstrated by Perkins and Mote \cite{PerkinsMote89} 
and again later by others investigating the ``chain fountain'' or closely related phenomena \cite{HannaKing11, HannaSantangelo12, Biggins14}.  

For lighter bodies and higher velocities, which have higher drag to weight ratios, drag plays a more noticeable role, affecting both shape and tension. 
Slow shape dynamics and wave behavior are still interesting features, constitute the remainder 
of the IPT problem \cite{IPT19}, and are on full display in Yeany's ``String Shooter/Launcher'' toy and science demonstration \cite{StringLauncher, YeanyYouTube}. 
An analytical solution for curvature and tension in terms of tangential angle 
was put forward by the astronomer 
Gregory in the 1940s \cite{Gregory49}, along with a brief report of some experiments.\footnote{\ldots not long before retiring from science to pursue interests in parapsychology \cite{GregoryObit1, GregoryObit2}.} 
Gregory cites no previous work--- later authors will reciprocate--- but indicates the following origin: ``It has long been known to engineers that a belt when driven at a high speed will occasionally be seen to stand up in the air in the form of a loop... 
The author's attention was drawn to this problem at the Admiralty when examining a proposal to use such a loop as a means of defence against aircraft.'' 
He expresses interest in using this problem to measure drag forces.  A similar sentiment, connected with the physics of fly fishing lines, drove the recent lariat chain experiments of Judd \cite{JuddLariat}. 
Gregory patches upper and lower curves together to create a loop. 
He realizes that at lower drag the tension is clearly not patchable, and that at higher drag, the curvature becomes singular at the patch point. He thus interprets his analytical solution as being only valid for intermediate values of drag, not recognizing the potential validity of singular high drag solutions.  
Analytical and experimental results appear in an independent body of work beginning in the 1960s 
\cite{Kurkin64, SvetlitskiiGabryuk66, Svetlitskii72inrussian, Svetlitskii72, Miroshnik72inrussian, SvetlitskiiMiroshnik73, MiroshnikKurkin76inrussian, Miroshnik01},\footnote{This body of work has gone uncited by later investigators of the string shooter, and the present authors only became aware of it during late-stage revisions.  These works also cite earlier and concurrent theoretical and experimental results by other authors, in references which we were unable to obtain at the time of writing.}
 inspired by antenna and radiator applications,  
and at times treating additional complications such as a normal drag force quadratic in velocity, a fluid-fluid interface, or a finite distance between launch and return points. 
Kurkin \cite{Kurkin64} expresses tension in terms of tangential angle, with arc length and thus any shape information remaining in quadrature. 
He reports experiments in vacuum demonstrating that increasing drag leads to narrower and higher loops, but does not explicitly recognize any transitions.  He proposes measuring the drag coefficient by using the balance of moments about the support. 
Svetlitskii and Gabryuk \cite{SvetlitskiiGabryuk66} find an implicit solution relating Cartesian coordinates and arc length, and show experimental shapes for moderate and high drag, noting that low drag solutions are not admissible. 
These shapes display features suggestive of bending stiffness. 
 As they do not solve for the curvature, they do not notice the second bifurcation. No theoretical curves are shown, and patching at the vertical is not discussed.  
Svetlitskii, Miroshnik, and Kurkin \cite{Svetlitskii72inrussian, Svetlitskii72} 
find an explicit Cartesian parameterization and expression for the tension, and favorably compare a high drag experimental shape with a solution, estimating the drag coefficient with Kurkin's method.  
They remain unaware of the second bifurcation, and do not provide a clear discussion of patching. 
Miroshnik \cite{Miroshnik72inrussian} expresses tension in terms of tangential angle. The string shooter is a special case of his more general solution, which features an additional purely normal drag force quadratic in velocity. He notes a singularity at the vertical and identifies the first bifurcation, showing high drag solutions while apparently unaware of the second bifurcation. 
The effects of angle and drag are shown. Examples include a vertically launched shape, which is impossible in the absence of normal drag--- it's not clear whether the author recognized this and verified the solution, but this point was made in \cite{Svetlitskii72inrussian}. 
Continuing work on this general solution, Svetlitskii and Miroshnik \cite{SvetlitskiiMiroshnik73} express curvature and tension in terms of tangential angle, thereby identifying both bifurcations, the corresponding behaviors at the vertical point, and the difference between moderate and high drag shapes.    
Miroshnik and Kurkin \cite{MiroshnikKurkin76inrussian} make use of known Cartesian solution forms for parameterization and tension and add to these an expression for arc length. By considering independent launch and ``return'' points, beyond simply two locations on a pulley, they explore tent-shaped and S-shaped solutions, and shape transitions under changes in drag and string length. 
Decades later, Miroshnik \cite{Miroshnik01} revisits and discusses many aspects of the problem, including the bifurcations and the vertical singularity, for similarly general boundary conditions. 
This important, yet overlooked, paper provides a Cartesian parameterization and expressions for tension, arc length, and curvature, and favorably compares experimental shapes with solutions for different drag regimes and launch angles, again estimating drag coefficients with Kurkin's method. The experimental shapes display features suggestive of bending stiffness, including a ``dolphin nose'' \cite{Judddolphin} in his figure 10.  Hanging low drag catenaries are also discussed.  Some examples of tension as a function of arc length are (parametrically) plotted. 
Miroshnik also suggests two additional ``modes'' for which the string's direction of motion is ascending through a vertical singularity, and dismisses them because they self-intersect. 
 However, such an ascending singularity is not possible because the arc length and tension are not regularized there, something this paper does not recognize, despite the observation in \cite{Svetlitskii72inrussian} that a vertical launch is not possible.

\section{Critique of recent literature}\label{critique}

Most of what is needed to understand the abstract string shooter problem is present in 
\cite{Gregory49, SvetlitskiiMiroshnik73, Miroshnik01, ChakrabartiHanna16}. In \cite{ChakrabartiHanna16}, the problem appears as a special case of a broader multiparameter family of solutions that include translations of the catenary at any angle with respect to gravity.  While anisotropic drag linear in velocity was considered, it was noted that for purely axially moving bodies, the solutions are more generally valid, as the drag is tangential and uniform regardless of its dependence on velocity. 
The solutions undergo two important bifurcations in tension and shape, whose implications for patching were given above in Section \ref{problemsummary}. 
The first of these, occurring at a drag to weight ratio of unity, allows the inertially-shifted tension to vanish at finite length on one side. 
The second of these, occurring at twice this ratio, leads to finite arc length curvature blowup. However, the
 authors of \cite{ChakrabartiHanna16}, including one of the authors of the present paper, made two errors.  First, and most importantly, they did not realize that the first bifurcation leads to an integrable singularity, allowing patching at finite length.  This led to strange results involving finite tensions despite infinite lengths, which are likely incorrect and require revisiting.\footnote{Contributing to this confusion were obstacles to continuing numerical solutions through vertical orientation, and the conceptualization of the solutions in terms of heteroclinic connections in ``phase portraits'' for angle and curvature; above the first bifurcation, the solutions reach the angle axis orthogonally instead of asymptotically approaching it, and are thus no longer unique there, but the corresponding change in appearance of the portrait is subtle.  
This error in interpreting the first bifurcation carried over into an aspect of the second, where it seemed that a special nonsingular solution existed right at the second bifurcation. As stated in \cite{ChakrabartiHanna16}, ``The existence of such a solution only under such restrictive conditions on speed is surprising, and this result will remain suspect until it is observed experimentally.''  This is now resolved by the recognition that the suspect result was simply wrong. }
Second, they made an incorrect statement to the effect that the change in tension from infinite to zero occurs at the leading rather than the trailing side of the catenary with respect to the axial flow \cite[(29)]{ChakrabartiHanna16}, when in fact the location of this change depends on the parameters of the problem; were the statement strictly correct, we would be unable to patch one leading upper and one trailing lower side as is necessary here to create our loops. 
Note that transitions at both sides are possible for the broader multiparameter family, but not for the present problem, which sits along the right edge of the diagram in \cite[figure 13]{ChakrabartiHanna16}. This asymmetry disallows the additional modes suggested in \cite{Miroshnik01}.

Bevan and Deane \cite{BevanDeane19} rediscover the solution 
and use it to describe the ``chain fountain'', a catenary arch with tangential drag forces. They don't realize that the drag force need not be linear. They do realize that low drag solutions cannot reach vertical orientation, which they do not need for their problem. They do note the significance of the first bifurcation, but, like Gregory before them, do not admit solutions above the second. 
Abello and co-workers \cite{Abello20} rediscover the Cartesian parameterization, which they interpret as being only valid above the first bifurcation.  

We conclude this section by addressing the relevant portions of two papers from 2019 that provide numerical solutions, and our interpretation of these in the light of previously published analytical solutions and bifurcations \cite{Gregory49, SvetlitskiiMiroshnik73, Miroshnik01, ChakrabartiHanna16} which we believe clarify some of the observations. The two groups appear to be unaware of available analytical solutions and their interpretations, including the bifurcation generating the high drag curvature singularity. 

Taberlet and co-workers \cite{Taberlet19} consider two paradigms, guided by experiments on heavy beaded chains and light strings draped over a pulley. 
For heavy bodies, they ignore drag, and further ignore the difficulties associated with vertical orientation by assuming such points sit on the pulley.   
They mistakenly say that the shifted tension at the bottom of the catenary is zero, which leaves nothing to balance gravity. 
For light bodies they ignore gravity--- the source of difficulty--- so that verticality is no longer forbidden. However, the only steady shapes with drag and without gravity are straight lines \cite{Svetlitskii72, ChakrabartiHanna16}.  
For general cases between these limits, numerical solutions to the steady-state balance equations are generated using an unspecified method to ``avoid the singularity''.  
A shape is shown for high drag, and data is presented for both moderate and high drag.
They also perform dynamic numerical simulations using 1000 beads and springs, for low through high drag values.  They infer the presence of the first bifurcation from the behavior of their solutions, but seem to be unaware of the second bifurcation, above which most of their results lie. It is not clear whether the low drag simulations reach steady state, or whether an effective ``bending'' due to discretization, or some dynamic effect, can allow closed loop solutions below the first bifurcation. 
   They report a numerical balance between the moments due to drag and weight forces, which is as expected if no moment is supplied at the support, and there is no net angular acceleration of the loop. 
The effects of drag and launch angle on shape and orientation are shown. 
An experimental image in their supplemental material shows a ``dolphin nose'' shape \cite{Judddolphin} which, along with other features visible throughout, is presumably due to bending stiffness. 

Daerr and co-workers \cite{Daerr19} numerically obtain a family of static string solutions for upper and lower parts of a loop for moderate and high drag, which they 
patch, guided by experimental shapes. 
Asymptotic behavior near the vertical is considered analytically; this expansion would not be possible at low drag due to the divergence of arc length at the vertical. 
Although bifurcations can be inferred from these expressions, they seem unaware of them, although they note qualitative differences in shape at high drag. They rediscover Kurkin's method for measuring drag coefficients. 
They employ two approaches to form loops. In one, they patch upper and lower curves at the vertical, empirically rediscovering Gregory's result of equal lengths. In the other, they create a large jump in angle 
by cutting off the small region of high curvature in the top part of the experimental curves (see \emph{e.g.} their figure 5c).  
They interpret this kink as something physical and measurable, saying the solution is obtained ``on condition that both the angle and the tension are allowed to be discontinuous'', but this is not physically admissible, as it requires an 
injection of force from some external source not present. 
We interpret the ``apparent angle discontinuity'' data reported in their figure 5a as an arbitrary choice of these authors, rather than a physical effect. 
They later invoke the possibility of nonlocal drag forces as the regularizer of this feature.

\section{Some misconceptions}\label{misconceptions}

After all the above, let us briefly note a few recurring misconceptions and some misleading language surrounding this problem that appear in the published scientific literature or academic discussions around the 
Tournament problem:\footnote{Regarding additional extensive discussions online on video sharing websites or other social media by professional science popularizers,  companies profiting from a redesign and renaming of the original invention, or enthusiastic members of the public, suffice it to say that ``someone is \emph{wrong} on the internet'' \cite{xkcd}.} 
\begin{itemize}
\item At high drag, above the often overlooked second bifurcation, there is no jump in angle at the vertical singularity, but the curvature blows up there. 
At low and moderate drag, the points of maximum curvature are instead at the top or bottom of the curves. 
The singularity always has to do with vertical orientation. 
\item The body is not ``lifted'' by drag forces; it is supported by the force supplied at the launch point.  
As noted in \cite{Kurkin64, Taberlet19, Daerr19}, there are no lift (normal drag) forces on a purely tangentially moving body, and the integral of a constant-magnitude tangentially-directed force around a closed loop is zero. 
However, as noted in \cite{Gregory49, Kurkin64, Miroshnik01, Taberlet19, Daerr19}, there is a moment about the support due to drag, which will balance the moment due to gravity. 
 But while these moments increase if the shape ``lifts'' to the side, they decrease again if the shape ``lifts'' farther upwards. 
While drag can ``raise'' the shape somewhat, a quite significant effect is achieved by changing the launch angle, as was shown for high drag in \cite{Miroshnik01} (see also \cite{Miroshnik72inrussian} for a case with additional normal flow), and will be shown below for low drag in Section \ref{catenaryloops}.\footnote{We note that recently published visual demonstrations of the necessity of air have been performed with boundary conditions 
either close to horizontal \cite{Taberlet19} 
or vertical \cite{Daerr19}. The latter is compared with a lifted-up state that does not appear to be steady; Gregory \cite{Gregory49} points out that a vertical boundary condition requires an infinite tension.}
Axial motion generates both inertial forces, which change the tension, and tangential drag forces, which change the tension and the shape. 
The former can, for example, stabilize very tall unclosed catenaries \cite{MillerYouTube} in the absence of air, or make heavy chains hover slowly \cite{Aitken1878, LariatChain}. 
Both this stabilization and the drag-induced bifurcations could resemble 
a sudden lifting action by the air at some threshold speed. 
\item Neither is the body ``launched'' into the air like a free projectile.  The driving wheels do work against the integrated tangential drag, as discussed below in Section \ref{global}.  The kinetic energy of a piece of material at the launch point does not determine the height of the loop, although higher velocities can stabilize taller catenaries by shifting the tension.  
\end{itemize}


\section{A few solutions and non-solutions}\label{solutions}

In this section, we present a selection of string and beam solutions to the axially moving loop problem posed above in Section \ref{problemsummary}. 
Some of these ``solutions'' are, like some previously published numerical results, not physically admissible, as they require external forces not present in the problem as posed. 
We do not examine the stability of solutions, and do not anticipate that this will display any interesting complexity, 
 noting simply that strings under compression ($\sigma < 0$) are unstable, but may be stabilized either by adding inertial forces--- independently of drag forces--- to bring them under tension, or by adding bending stiffness, such that compression is permitted.
Inertia alone is irrelevant to most aspects of the problem, serving merely to shift the tension by a constant, whereas drag forces can affect both the shape, and qualitative features of the tension profile.

\subsection{Catenary loops}\label{catenaryloops}

The string equations for an inextensible curve $\bm{x}(s,t)$ parameterized by arc length $s$ and time $t$, with uniform mass density $\rho$, in the presence of gravity $-g \bm{\hat{e}}_2$ and a drag force $\bm{D}$, are the balance of linear momentum
\begin{equation}\label{eq:eom0}
    d_s(\sigma d_s\bm{x}) - \rho g \bm{\hat{e}}_2 +\bm{D} = \rho d_t^2\bm{x} \, ,
\end{equation}
and the inextensibility constraint 
\begin{equation}
	d_s\bm{x}\cdot d_s\bm{x} = 1 \, ,\label{inextensibility}
\end{equation} 
enforced by a multiplier $\sigma(s, t)$ which we identify as the (bare, unshifted) tension.  There is no need to independently consider angular momentum balance.  
Noting that the curve will be coplanar with gravity, we further define a $\bm{\hat{e}}_1$ axis perpendicular to gravity, and a tangential angle $\theta(s,t)$ describing the orientation of the curve's 
unit tangent $\bm{\hat{e}}_t = d_s\bm{x} = \bm{\hat{e}}_1\cos\theta + \bm{\hat{e}}_2\sin\theta$  
and unit normal $\bm{\hat{e}}_n = -\bm{\hat{e}}_1\sin\theta + \bm{\hat{e}}_2\cos\theta$, 
so that $d_s\bm{\hat{e}}_t = d_s\theta\bm{\hat{e}}_n$ and $d_s\bm{\hat{e}}_n = -d_s\theta\bm{\hat{e}}_t$. 
The out of plane unit vector is $\bm{\hat{e}}_3$, and both Cartesian and curve frames are right-handed. 
 
We consider only shape equilibria, in which material flows along a steady shape with a uniform and constant tangential speed $v$.  
This implies that $d_t\bm{x} = v \bm{\hat{e}}_t $, $d_t^2\bm{x} 
= v^2 d_s\theta \bm{\hat{e}}_n$, and $\bm{D} = -D(v) \bm{\hat{e}}_t$. 
The form of $D(v)$ may remain unspecified. Physically, it will be some nonnegative, presumably monotonic, function of $v$, and is often assumed quadratic. We hereafter stop writing the explicit dependence of $D$ on $v$. 
Further nondimensionalizing $\bm{x}$ and $s$ by the total string length $\ell$, $D$ and $\sigma$ by $\rho g \ell$, and $d_t^2\bm{x}$ by $g$, 
 the equilibrium equation takes the simple form
 \begin{equation}\label{eq:eom1}
    d_s\left[\left(\sigma -  v^2\right) \bm{\hat{e}}_t\right] = \bm{\hat{e}}_2 + D \bm{\hat{e}}_t \, ,
\end{equation}
with normal and tangential projections
\begin{align}
	(\sigma-v^2) d_s\theta &= \cos\theta \, , \label{eq:normproj1}\\
	d_s\sigma &= \sin\theta + D \, , \label{eq:tangproj1}
\end{align}
where we have used the fact that $d_s(\sigma-v^2) = d_s\sigma$. 
The two fields $\sigma$ and $\theta$ depend on $s$, while the parameters $v$ and $D$ are uniform.  A space-fixed boundary or patching point is non-material, so a jump condition \cite{OReilly17book} 
obtains there: 
\begin{align}
	{\bf{R}} + \llbracket (\sigma -  v^2 ) \bm{\hat{e}}_t \rrbracket = \bm{0} \, ,
\end{align}
where ${\bf{R}}$ is a singular source of linear momentum, as may be provided only at the support point.  In other words, when patching two solutions together, a jump in the quantity $\left(\sigma -  v^2\right)\bm{\hat{e}}_t$ requires an external force. It follows that the shifted tension $\sigma -  v^2$ must vanish at an unsupported (${\bf{R}}=\bm{0}$) kink.\footnote{The exceptional case of a cusp, where the orientation of $\bm{\hat{e}}_t$ jumps by $\pi$, will not be considered.}

Generic solutions without drag $(D=0)$, including the static $(v=0)$ catenaries, only asymptotically approach $\theta \rightarrow \pm \frac{\pi}{2}$ as the  arc length $s \rightarrow \pm \infty$.  Concurrently, the shifted tension $\sigma - v^2 \rightarrow \pm \infty$, while the curvature $d_s\theta \rightarrow 0$, as does their product balancing the normal contribution of gravity $\cos\theta \rightarrow 0$. 
A finite loop cannot be closed without a kink and associated external force, and the diverging tension is a significant obstacle to patching a hanging tensile catenary to a compressive catenary arch. 

Solutions with drag can be obtained following the derivation in \cite{ChakrabartiHanna16}, as detailed in \ref{derivationcatenary} of the present paper. 
The tension is eliminated to form a single equation for $\theta$ with one parameter $D$, which is then integrated to obtain the results:
\begin{align}
    d_s\theta &= \dfrac{C \cos^2\theta}{\left(\sec\theta+\tan\theta\right)^{D}} \,, \label{eq:curvature} \\
    \sigma - v^2 &= \dfrac{\left(\sec\theta+\tan\theta\right)^{D}}{C\cos\theta } \,, \label{eq:tension}
\end{align}
which may be compared with \cite[equations 30-31]{ChakrabartiHanna16}. 
 The integration constant $C$ relates the total length and subtended angle of a segment of curve. 
 These solutions, whose singularities at the vertical depend on the size of $D$, are valid as written for ranges of $\theta$ such that the parenthetical quantity is positive. 
We will typically be working with clockwise-parameterized upper branches, where $\theta$ starts near $\frac{\pi}{2}$ and decreases to near $-\frac{\pi}{2}$, and $C$ is negative; lower branches are constructed from these by symmetry operations. 
This and other issues related to numerical implementation and patching of upper and lower branch solutions to form a closed loop are detailed in \ref{patchingcatenary}-\ref{gregory}. 

Solutions of the shifted tension \eqref{eq:tension} and curvature \eqref{eq:curvature} equations respectively undergo distinct bifurcations at $D=1$ and $D=2$, related to the nature of singularities at vertical orientation.\footnote{\label{footnotederivatives}Some of the behavior of solutions can also be inferred from an expansion of (\ref{eq:normproj1}-\ref{eq:tangproj1}) around the vertical $\theta=-\frac{\pi}{2}$ if the arc length is finite there, requiring $D>1$. Denoting the deviation in angle and arc length from this point respectively as $\Theta$ and $S$, one obtains an approximate dynamical system $d_S \Theta = \Theta / (D-1)S$ with solutions $\Theta \sim S^{1/(D-1)}$, the latter also obtained by \cite{Daerr19}.  
For $1<D<2$, the exponent is greater than unity, and both $\Theta \rightarrow 0$ and $d_S\Theta \rightarrow 0$ as $S\rightarrow 0$, with non-unique solutions at $S=0$, a feature missed in \cite{ChakrabartiHanna16}. For $D>2$, the exponent is less than unity but still positive, and $\Theta \rightarrow 0$ and $d_S\Theta \rightarrow \infty$ as $S\rightarrow 0$. 
 This analysis suggests the presence of an infinite sequence of bifurcations through which each higher derivative of $\theta$ transitions from vanishing to blowing up at the vertical; for example, the second and first derivatives of curvature $d_S^3\Theta$ and $d_S^2\Theta$ should respectively undergo transitions at $D=4/3$ and $D=3/2$.  
The tendency of a quantity to vanish or blow up may become weak near bifurcations, as reflected in the relevant exponents. 
}
Referring to the regimes discussed in Section \ref{problemsummary}, 
\begin{itemize}
	\item For low drag ($D < 1$), the singularities are qualitatively like those of a catenary without drag forces.
	\item For moderate drag ($1 < D < 2$), at one vertical point $\sigma - v^2 \rightarrow 0$ at finite $s$.
	\item For high drag ($2 < D$), at one vertical point  $| d_s\theta | \rightarrow \infty$ at finite $s$. 
\end{itemize}
As also discussed in \cite{ChakrabartiHanna16}, special cases occur at the two bifurcations.  When drag and weight are equal ($D=1$), the shifted tension takes on a finite value at the vertical, 
 and when drag is twice the weight
($D = 2$), the curvature takes on a finite value at the vertical. 
The physical meaning of these special cases remains unclear, and neither has been experimentally realized.

For the problem as defined, it is only necessary to deal with one free vertical singularity, at $\theta = -\frac{\pi}{2}$.  On the other side, jumps in angle and tension can be absorbed into conditions at the support point. 
There appears to be no way to have these solutions pass through the vertical in two locations, and thus no way to fully close an unsupported (no kink) loop of finite length. 
With a single support, low drag loop solutions cannot be constructed, but moderate and high drag solutions can be formed by patching upper and lower branches at an unsupported vertical, although the high drag branches terminate with infinite curvature. 

\begin{figure}[h!]
     \centering
     \includegraphics[width=0.95\textwidth]{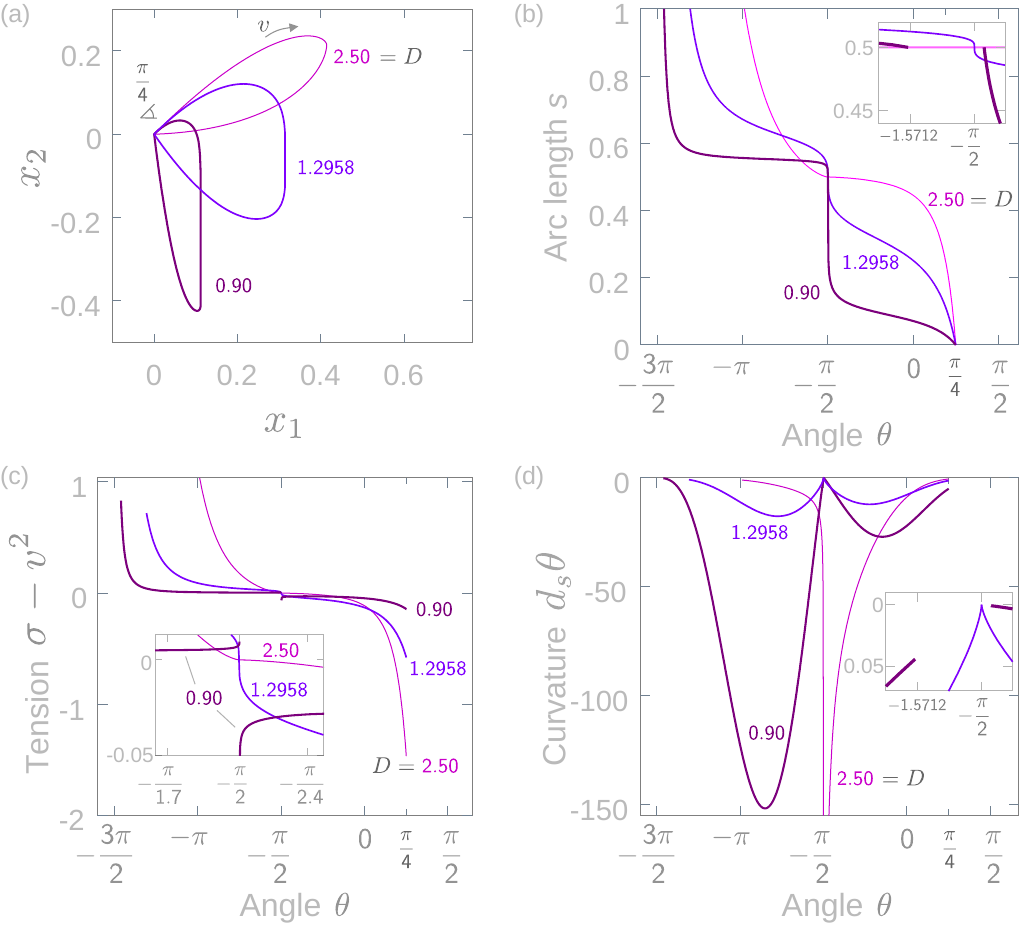}
     \caption{Catenary loops with low, moderate, and high drag ($D = 0.90, 1.2958, 2.50$) and launch angle $\theta(0) = \frac{\pi}{4}$. (a) Shapes. (b) Arc length $s$, (c) shifted tension $\sigma - v^2$, and (d) curvature $d_s\theta$, as functions of angle $\theta$. 
     The moderate drag curvature cusp in (d) reaches the axis at $\theta=-\tfrac{\pi}{2}$ by orthogonal approach.}
        \label{strings45}
\end{figure}

In four figures, two in this subsection and two in the next, we display example shapes with launch/supply point at the origin, as well as the corresponding behavior of the arc length, shifted tension, and curvature, in terms of  the angle. The direction of motion $v$ and increasing arc length $s$ is clockwise.  

Figure \ref{strings45} presents one example each of low, moderate, and high drag catenaries, launched at an angle of $\frac{\pi}{4}$.  
The low drag string is not a proper solution, as it does not satisfy linear momentum balance without the inclusion of a second  force at the vertical point, in addition to the allowed support force at the origin. The inset in (c) reveals the tendency for the tension to diverge at the vertical ($\theta=-\frac{\pi}{2}$). This divergence, and the approach of curvature to zero, are truncated by the presence of a kink, revealed as a gap in angle in the insets in (b) and (d). 
This gap can only be closed as the length of the string diverges; note that arc length is shown rescaled by this total length. 
The tension divergence is eliminated at moderate drag, for which subtle but important changes in shape are present, including the curvature curve in (d) reaching the axis at $\theta=-\frac{\pi}{2}$ by orthogonal approach. Further perspective on the derivatives of the $\theta$ curve in (b), which are all zero or infinite at the vertical, can be found in the footnote\footref{footnotederivatives}. 
A qualitative change in shape is apparent at high drag, for which the curvature diverges at the vertical, although there is no clear visual signature of the singularity itself.   

\clearpage

\begin{figure}[h!]
     \centering
     \includegraphics[width=0.95\textwidth]{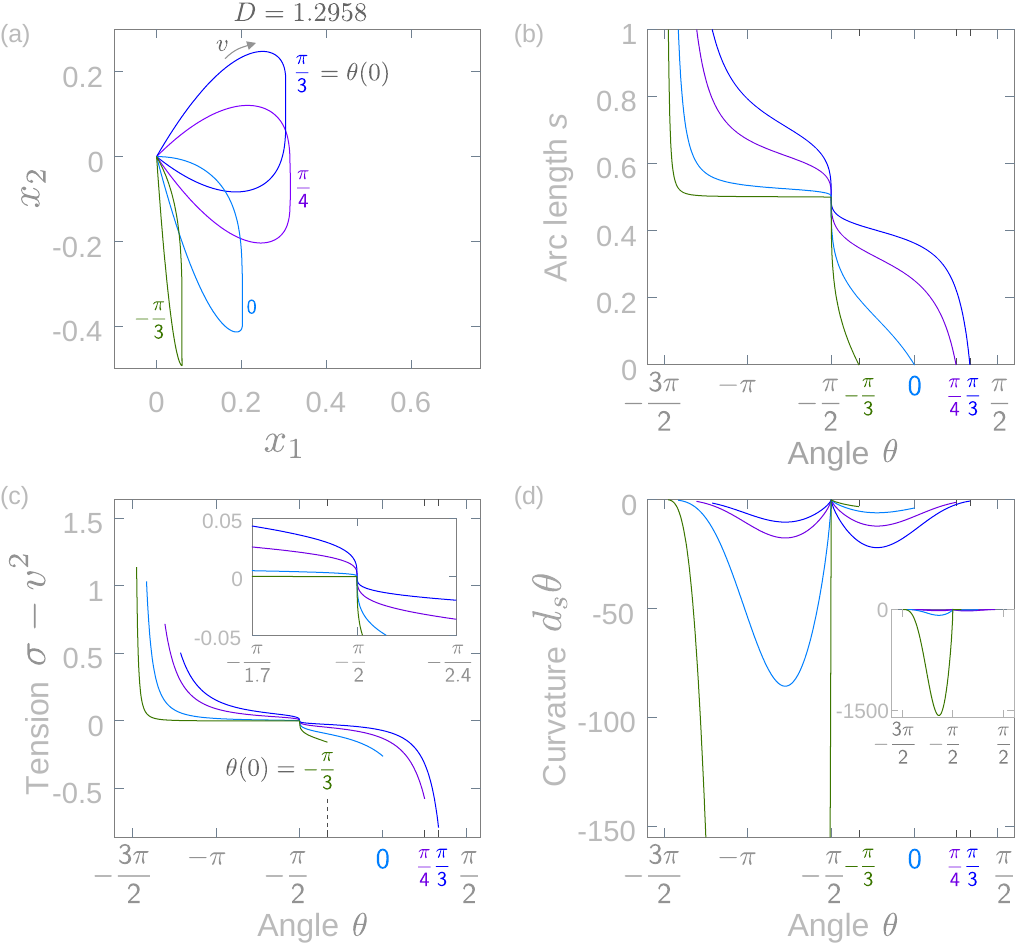}
    \caption{Catenary loops with moderate drag $D = 1.2958$ and launch angles $\theta(0) = -\frac{\pi}{3}, 0, \frac{\pi}{4}, \frac{\pi}{3}$. 
    (a) Shapes. (b) Arc length $s$, (c) shifted tension $\sigma - v^2$, and (d) curvature $d_s\theta$, as functions of angle $\theta$. Curves in (d) intersect on the axis at $\theta=-\tfrac{\pi}{2}$. }
        \label{stringsD}
\end{figure}

Figure \ref{stringsD} presents moderate drag catenaries with different launch angles.  It can be seen that varying the launch angle ``lifts'' the string as much as drag does. 
The solutions are non-unique, as the curves in (d) intersect on the axis at $\theta=-\frac{\pi}{2}$. 
However, this seems unlikely to correspond to any possibility of coexisting states in the string shooter problem as typically posed. One might fix the drag $D$ by setting the speed, and impose some launch angle.  
The other features of the response appear to be unique. For example, the return angle would be determined by Gregory's \cite{Gregory49} transcendental relation \eqref{eq:thetareturn}, reproduced in our \ref{gregory}.
  

\subsection{Regularization by bending stiffness}\label{bendingstiffness}

It is easy enough to physically hang a static loop from a support, or launch a slow-moving one, yet the physics of a catenary forbids such solutions.  
The most straightforward resolution of this issue comes from the addition of a bending stiffness, which allows for nonvertical inflectional, and vertical, points. Such points are clearly present in experiments, including for example in
\cite[figure 3]{SvetlitskiiGabryuk66}, \cite[figures 10 and 11]{Miroshnik01}, \cite[figure 1c]{Daerr19}, and \cite[figure 2a]{Taberlet19}, the latter reminiscent of \cite[figure 4d]{WatsonWang81} and \cite[plate III figure 3]{Aitken1878}.  
The resulting problem is that of the heavy \emph{elastica} or stiff catenary \cite{Wang86}.  In the static limit, the physical situation is that of holding an elastic loop 
 at one point, as between two fingers that supply a force and moment. We will also still allow for the presence of a kink. 

The regularization required at low drag is not a smoothing of the sharp angle jump with a high curvature region, but rather has to do with the divergence of the arc length and tension as the vertical is approached. The important new forces that balance the tension-curvature term at verticality will be associated with the second derivative of curvature, rather than curvature itself. 
We do not explore stiff solutions at high drag, which would presumably regularize, in a more traditional form, the singularities in curvature of such solutions. 

We augment the contact force to be $\sigma d_s \bm{x} - Ed_s^3 \bm{x}$, where $E$ is a uniform, dimensionless bending stiffness. Recall that $\bm{\hat{e}}_t  = d_s \bm{x}$. This modifies the linear momentum balance for equilibria \eqref{eq:eom1}, 
\begin{align}\label{eq:elasticaeom}
    d_s\left[\left(\sigma -  v^2\right) \bm{\hat{e}}_t - Ed_s^3 \bm{x} \right] = \bm{\hat{e}}_2 + D \bm{\hat{e}}_t \, , 
\end{align}
and its normal and tangential projections (\ref{eq:normproj1}-\ref{eq:tangproj1}), 
\begin{align}
    \left[ \sigma - v^2 + E \left(d_s\theta \right)^2 \right] d_s\theta - Ed_s^3\theta &= \cos\theta \, , \label{eq:elastican}\\    
  d_s\sigma + 3E d_s\theta d_s^2\theta &= \sin\theta + D \, . \label{eq:elasticat}
\end{align}
We retain the inextensibility constraint \eqref{inextensibility}.  
All this is sufficient to solve for $\sigma$ and $\theta$. 
Defining $\bm{n} \equiv \sigma\bm{\hat{e}}_t - Ed_s^3 \bm{x}$ and $\bm{m} \equiv E d_s\theta \bm{\hat{e}}_3$, 
the angular momentum balance
\begin{align}
	d_s \bm{m} + \bm{\hat{e}}_t \times \bm{n} = \bm{0} \, , \label{eq:elasticaangular}
\end{align}
is potentially redundant, as it provides the same information as the normal projection of the linear momentum balance. However, we use it in our numerical formulation (detailed in \ref{numelastica}), and we will refer to it later in a discussion of global balances (Section \ref{global}). 
Defining the quantity appearing in the derivative in \eqref{eq:elasticaeom} as $\bm{f} \equiv \bm{n} - v^2\bm{\hat{e}}_t$, the shifted tension is now $\bm{f}\cdot\bm{\hat{e}}_t = \sigma - v^2 + E (d_s\theta)^2$. 
Jump conditions \cite{OReilly17book} for these equilibrium configurations are
${\bf{R}} + \llbracket \bm{f} \rrbracket = \bm{0}$
and ${\bf{M}} + \llbracket \bm{m} \rrbracket = \bm{0}$, 
where the latter condition, allowing for the presence of a singular source of angular momentum ${\bf{M}}$ at the support, is identical to a static boundary condition for this beam model without rotary inertia.  
The only jump is at the support point, as there is no longer a need for patching at the vertical.  
It is apparent from \eqref{eq:elastican} that the $d_s^3\theta$ term, the second derivative of curvature, is what will balance the product of shifted tension and curvature at the vertical, where $\cos\theta=0$. 
 
\begin{figure}[h!]
      \centering
      \includegraphics[width=0.95\textwidth]{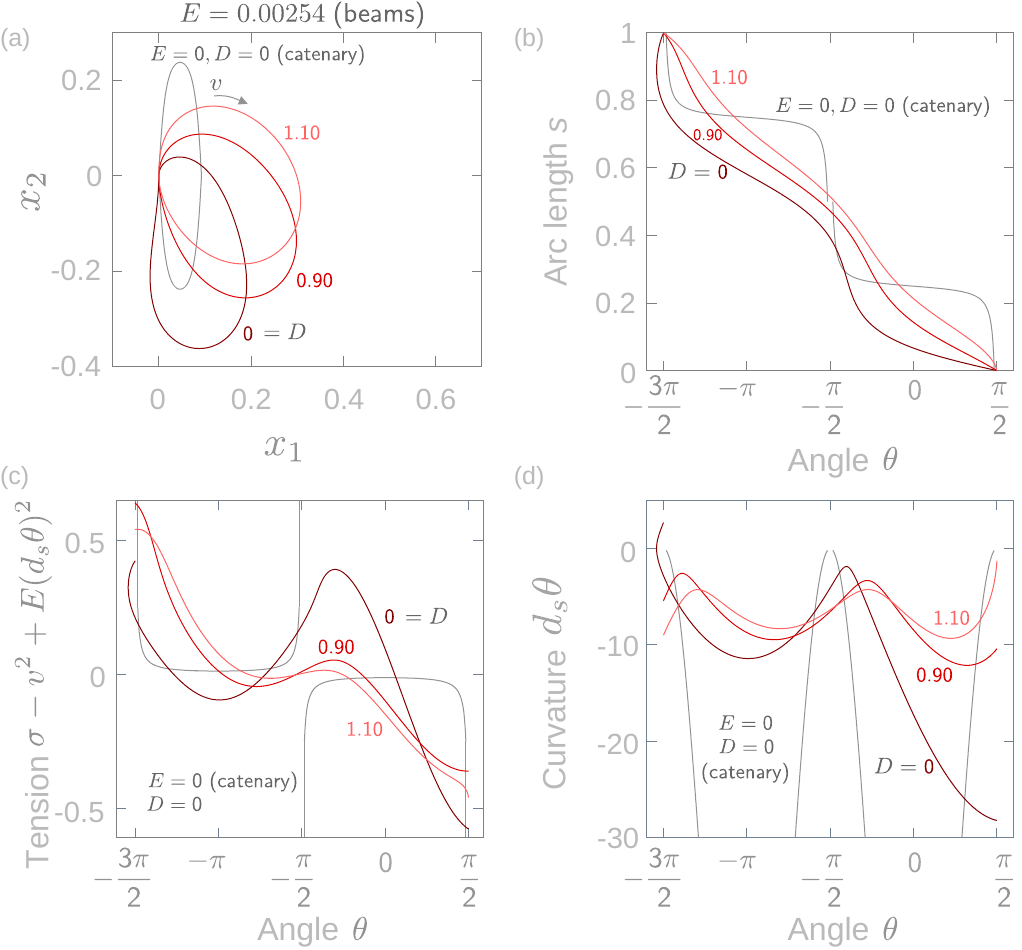}
     \caption{Stiff catenary loops with bending stiffness $E = 0.00254$ and zero, low, and moderate drag ($D = 0, 0.90, 1.10$) and a kink-free vertical orientation at the support (launch angle $\theta(0)=\frac{\pi}{2}$, return angle $\theta(1)=-\frac{3\pi}{2}$), alongside a classical static non-stiff ($D=0, E=0$) catenary with an angle cutoff of $\pm0.05$. 
     (a) Shapes. (b) Arc length $s$, (c) shifted tension $\sigma - v^2 + E\left(d_s\theta\right)^2$, and (d) curvature $d_s\theta$, as functions of angle $\theta$. The curvature of the catenary is approximately $-79.933$ at $-\pi$ and $0$, and approximately $-0.200$ at the patching point. } 
        \label{egg90}
\end{figure}

Figure \ref{egg90} presents stiff catenaries (beams) with zero, low, and moderate drag, and a kink-free vertical orientation at the support, alongside a static non-stiff catenary (string) with a small angle cutoff for qualitative comparison. 
 The catenary fails to satisfy momentum balance by a large amount. Its angle gap, diverging tension, and nonzero curvature at the patching point near the vertical $\theta = -\frac{\pi}{2}$ are clearly visible across (b), (c), and (d).  
 These features are absent in the presence of bending stiffness. No obvious signature of the $D=1$ bifurcation remains. Both the shifted tension and the curvature are nonzero at the vertical. 

\clearpage

\begin{figure}[h!] 
     \centering
     \includegraphics[width=0.95\textwidth]{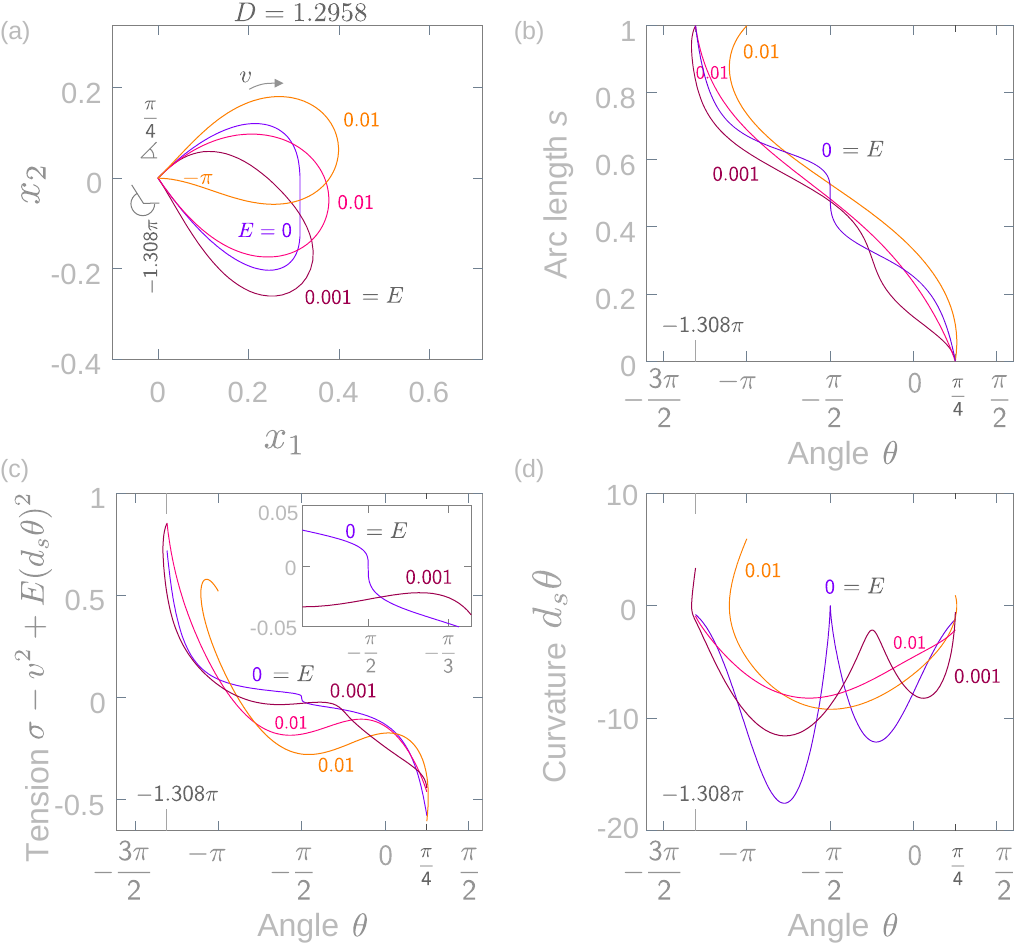}
     \caption{Stiff catenary loops with moderate drag $D = 1.2958$, launch angle $\theta(0) = \frac{\pi}{4}$, and different different return angles or stiffnesses ($E=0.01$, $\theta(1) = -1.308\pi$), ($E=0.001$, $\theta(1) = -1.308\pi$), ($E=0.01$, $\theta(1) = -\pi$), alongside a non-stiff ($E=0$) catenary with the same drag and launch angle, and the same return angle ($\theta(1) = -1.308\pi$) as two of the stiff catenaries. 
          (a) Shapes. (b) Arc length $s$, (c) shifted tension $\sigma - v^2 + E\left(d_s\theta\right)^2$, and (d) curvature $d_s\theta$, as functions of angle $\theta$.}
        \label{egg45}
\end{figure}

Figure \ref{egg45} presents stiff catenaries with moderate drag, launch angle $\frac{\pi}{4}$, and different return angles or stiffnesses, 
alongside a non-stiff catenary with the same drag, launch angle, and return angle as two of the beams. 
Allowing a kink in an axially moving elastic beam can instead be interpreted as material being fed in and out at the launch point, rather than a driven closed loop.  
Note that both inflectional and non-inflectional \emph{elastica} are represented, depending on boundary conditions. 

It is clear that even a small amount of bending stiffness leads to drastic changes in shape, creating the eggplant or slumping balloon shapes of Figures \ref{egg90}-\ref{egg45}.  
Figure \ref{egg45}(a) can be qualitatively compared with the moderate drag experimental shapes from \cite[figure 3]{SvetlitskiiGabryuk66} and \cite[figure 11]{Miroshnik01}, although these works do not consider bending.

\subsection{Global balances}\label{global} 

Integrating the linear \eqref{eq:elasticaeom} and angular \eqref{eq:elasticaangular} momentum balances over the loop provides the total force and moment supplied by the support at the boundary: 
\begin{align}
    \bm{f}(1) - \bm{f}(0) &= \int_0^1\!\! \left(\bm{\hat{e}}_2 + D\bm{\hat{e}}_t \right) ds = \bm{\hat{e}}_2 \, ,\label{eq:globalf} \\
    \bm{m}(1) - \bm{m}(0) &= \int_0^1\!\! \bm{n}\times\bm{\hat{e}}_t \, ds 
    = E \left[ d_s\theta(1) - d_s\theta(0) \right]\bm{\hat{e}}_3
 \label{eq:globalm} \, ,
\end{align} 
the latter also following directly from the definition of $\bm{m}$. 
These two quantities can be identified as ${\bf{R}}$ and ${\bf{M}}$ at the support point, where a jump $\llbracket q \rrbracket = q(0)-q(1)$. 
The force is purely vertical, balancing gravity; the drag does not contribute. 
In the string limit ($E=0$), it is not possible to apply a moment, and as noted in \cite{Gregory49, Taberlet19, Daerr19}, the integral moments about the support due to drag and gravity balance each other; we can also infer that the moment about the center of mass due to drag balances that due to the support force.

It is illustrative to integrate the tangential projection of the linear momentum balance for axially moving equilibria \eqref{eq:elasticaeom}, not simplified into \eqref{eq:elasticat}, but rather rearranged into 
 the form of a pseudomomentum balance \cite{MaddocksDichmann94, OReilly17book, SinghHanna21} 
(see also \cite{Broer70, Healey96} for strings): 
\begin{align}
	d_s\left[  \sigma - v^2 + \tfrac{3}{2}E \left(d_s\theta \right)^2 - \bm{\hat{e}}_2\cdot\bm{x}  \right] = D \, , \label{eq:pseudo}
\end{align}
which integrates over the closed unit loop to
\begin{align}
   \left[  \sigma + \tfrac{3}{2}E \left(d_s\theta \right)^2  \right] \Big|_0^1  = D \, .  
     \label{eq:pseudo2}
\end{align}
We may thus identify a singular source of pseudomomentum at the support, required to drive the body through the resistive fluid medium.  
Given the close connection between the body tangents and the velocity in inextensible axially moving equilibria, it is to be expected that the pseudomomentum balance would express similar information as an energy balance. 

\section{Open questions}

While a small amount of bending offers the simplest physically plausible regularization of the vertical singularity, the resulting numerical problem is quite stiff, making comparison with experiment difficult. 
This suggests the development of a matched asymptotic analysis, likely to be quite different from prior analyses for heavy \emph{elastica} in which the boundary layer is due to clamping \cite{Plunkett67, StumpvanderHeijden00, Wolfe93}. 
We expect the ``dolphin nose'' \cite{Judddolphin}, visible in experiments, to emerge at lower bending stiffnesses than those examined here.

It is also worth considering unsteady shapes and their associated inertial and drag forces, including normal drag.  No one has experimentally, or even numerically, clearly demonstrated perfectly steady motion; in experiments, there are pulsatile motions repeatedly generated by the interaction between the driving wheel(s) and the body.  It could be that some dynamic effect, rather than bending, is responsible for creating almost-steady configurations at low drag. 


The vertical singularity, at which derivatives of angle either vanish or blow up, would be interesting to explore further, particularly in the moderate drag range where all the transitions of the derivatives occur, as discussed in a previous footnote.\footref{footnotederivatives}  The sequence of shape bifurcations is likely to be visually subtle, but may lead to numerical issues. 
We are not immediately aware of another simple physical problem generating a toy mathematical problem with this type of singular behavior. Whether any transitions remain or arise upon the addition of bending is unknown.

\section{Summary and Conclusions}

The effects of drag on string shooter solutions can be briefly summarized as follows. 
For a stationary shape, drag forces do not contribute to the normal force balance, but they change the equilibrium shape and tension.  
At low values, they do not change the qualitative features of the catenary, in particular its asymptotic approach to verticality, and the blowup of tension of opposite signs on upper and lower branches.  Additional physics, such as bending elasticity, is necessary to construct loop solutions. 
At moderate values, the inertially-shifted tension vanishes along with the curvature at verticality at finite arc length, allowing patching of loop solutions.
At high values, the curvature blows up at verticality, with patching still possible.  

We have corrected several errors in recent works on axially moving string loops coplanar with gravity.  
These entail misunderstandings of the role of drag forces, bifurcations, the nature of the vertical singularity, and admissibility of certain solutions. 
Several example solutions were constructed, either by implementing previously derived analytical results, or numerically adding bending stiffness. 
We have also attempted to place the problem in context alongside several other fundamental problems in flexible structure mechanics.

\section*{Acknowledgments}
We thank B. Chakrabarti for discussions, T. J. Healey, J. N. Papadopoulos, and N. C. Perkins for answering historical questions about their work, and especially thank N. Tuck and B. Yeany for providing detailed histories of their respective inventions. 
JAH thanks G. Judd and J. Miller for extensive stimulating correspondence over the last twelve years, including sharing of unpublished experimental results.

\appendix

\section{Practical details}\label{practical}

\subsection{Derivation of catenary solutions}\label{derivationcatenary}

The normal and tangential projections (\ref{eq:normproj1}-\ref{eq:tangproj1}) of the equilibrium equation for an axially moving catenary can be manipulated into the expression
\begin{align}
   \dfrac{d_s^2\theta}{d_s\theta} &= - \left(2\tan\theta + D \sec\theta\right) d_s\theta \, , \nonumber
\end{align}
and then integrated to obtain the curvature $d_s\theta$ and, from that, the tension $\sigma$,
\begin{align}
    d_s\theta 
    &=  \dfrac{C \cos^2\theta}{\left(\sec\theta+\tan\theta\right)^{D}}
    = 4 C \tan^{2-D}\left(\tfrac{\pi}{4}+\tfrac{\theta}{2}\right) \left[1 + \tan^2 \left(\tfrac{\pi}{4}+\tfrac{\theta}{2}\right)\right]^{-2} \, , \label{eq:curvatureapp} \\
    \sigma - v^2 &=  \dfrac{\left(\sec\theta+\tan\theta\right)^{D}}{C\cos\theta }
        = \frac{1}{2C}\left[ \tan^{D+1}\left(\tfrac{\pi}{4}+\tfrac{\theta}{2}\right) + \tan^{D-1}\left(\tfrac{\pi}{4}+\tfrac{\theta}{2}\right) \right] \,, \label{eq:tensionapp}
\end{align}
where $C$ is a constant of integration. For general values of $D$, these expressions are valid as written for the range $-\frac{\pi}{2} < \theta < \frac{\pi}{2}$, corresponding in our work to an upper branch of clockwise-travelling string. 
The compact expressions in the middle are preferred, but the rewritten expressions on the right are intended for further comparison with the work of Gregory \cite{Gregory49}, and were obtained using the identities $\sec\theta + \tan\theta = \tan\left(\frac{\pi}{4}+\frac{\theta}{2}\right)$ and $\sec\theta - \tan\theta = \cot\left(\frac{\pi}{4}+\frac{\theta}{2}\right)$.

The shape of the curve is determined by the curvature, 
\begin{align}
    s\, \big|_{\theta_1}^{\theta_2} \!&=\!\! \int_{\theta_1}^{\theta_2}\!\! \dfrac{1}{d_s\theta} \,d\theta
    = \dfrac{\left(D - \sin\theta\right)\left(\sec\theta + \tan\theta\right)^{D} }{C \left(D^2 - 1\right) \cos\theta} \Bigg|_{\theta_1}^{\theta_2} \!\!
    = \frac{1}{2 C}\left[\dfrac{\tan^{D+1}\left(\frac{\pi}{4}+\frac{\theta}{2}\right)}{D+1} + \dfrac{\tan^{D-1}\left(\frac{\pi}{4}+\frac{\theta}{2}\right)}{D-1}\right] \!\Bigg|_{\theta_1}^{\theta_2}    \label{eq:s}    \\
    x\, \big|_{\theta_1}^{\theta_2} \!&=\!\! \int_{\theta_1}^{\theta_2}\! \dfrac{\cos\theta }{d_s\theta} \,d\theta
    = \dfrac{\left(\sec\theta + \tan\theta\right)^D}{C D} \Bigg|_{\theta_1}^{\theta_2} \!\!
    = \dfrac{\tan^D\left(\frac{\pi}{4}+\frac{\theta}{2}\right)}{C D}  \Bigg|_{\theta_1}^{\theta_2} \label{eq:x}  \\
    y\, \big|_{\theta_1}^{\theta_2} \!&=\!\! \int_{\theta_1}^{\theta_2}\! \dfrac{\sin\theta }{d_s\theta} \,d\theta
    = \dfrac{\left(D\sin\theta - 1\right)\left(\sec\theta + \tan\theta\right)^{D} }{C \left(D^2 - 1\right) \cos\theta} \Bigg|_{\theta_1}^{\theta_2} \!\!
    = \frac{1}{2 C}\left[\dfrac{\tan^{D+1}\left(\frac{\pi}{4}+\frac{\theta}{2}\right)}{D+1} - \dfrac{\tan^{D-1}\left(\frac{\pi}{4}+\frac{\theta}{2}\right)}{D-1} \right] \!\Bigg|_{\theta_1}^{\theta_2} \label{eq:y}
\end{align} 
where, again, the range $-\frac{\pi}{2} < \theta < \frac{\pi}{2}$ is implied.

\subsection{Patching of catenary solutions}\label{patchingcatenary}

Implementation of the above results involves solving and matching two boundary value problems.  This can be numerically stiff, particularly near the first bifurcation where the shapes take a rather sharp turn near the patching vertical. 
Below the first bifurcation, the asymptotic approach to verticality necessitates use of a cutoff parameter $\epsilon$ for the angle there. 
We must solve separately for the top ($\frac{\pi}{2}$ to $-\frac{\pi}{2}$) and bottom ($-\frac{\pi}{2}$ to $-\frac{3\pi}{2}$) portions of the loop, which each have their own constant of integration $C$. 
We generate the bottom portion by creating a second top portion and reassigning quantities. 
The procedure is as follows for a loop of length $\ell = \ell_\text{up} + \ell_\text{down} =1$.
The top portion of the loop, 
with $0 \le s_\text{up} \le \ell_\text{up}$ 
subtending $\theta_\text{launch} \ge \theta_\text{up} \ge -\tfrac{\pi}{2}+\epsilon_\text{up}$ (here $s$ is increasing in the clockwise direction, with $\theta$ decreasing) is generated for a chosen launch angle $\theta_\text{launch}$ and length $\ell_\text{up}$.  
This first involves determining $C_\text{up}$ and $\epsilon_\text{up}$, by using expression \eqref{eq:s} and the identification $s\left(-\tfrac{\pi}{2}+\epsilon_\text{up}\right)  = \ell_\text{up}$ 
to numerically minimize the cutoff $\epsilon_\text{up}$ (for $D > 1$ where no cutoff is required, this procedure effectively leads to $\epsilon_\text{up} = 0$ within numerical error). 
Then the positions $x_\text{up}$ and $y_\text{up}$ are obtained from expressions \eqref{eq:x} and \eqref{eq:y}.
Next, we follow a similar procedure to construct what will become the bottom portion of the loop.
First, we find another ``top'' portion,  
with $0 \le s_\text{down} \le \ell_\text{down}$ 
subtending $-\pi - \theta_\text{return} \ge \theta_\text{down} \ge -\tfrac{\pi}{2} + \epsilon_\text{down}$. 
We determine $C_\text{down}$ and $\epsilon_\text{down}$, this time by identifying arc length, $x$ positions, and reflected $y$ positions at the patching point: $s_\text{down}\left(-\tfrac{\pi}{2}+\epsilon_\text{down}\right) = \ell - \ell_\text{up} = \ell_\text{down}$, $x_\text{down}\left(-\tfrac{\pi}{2}+\epsilon_\text{down}\right) = x_\text{up}\left(-\tfrac{\pi}{2}+\epsilon_\text{up}\right)$, and $y_\text{down}\left(-\tfrac{\pi}{2}+\epsilon_\text{down}\right) = -y_\text{up}\left(-\tfrac{\pi}{2}+\epsilon_\text{up}\right)$. Note that this patching point can be in the first or fourth quadrant, above or below the launch point. 
Finally, we shift and flip the curve, reassigning $s \rightarrow \ell - s$ and $\theta \rightarrow -\pi - \theta$ to create a bottom portion, with $\ell - \ell_\text{down} \le s_\text{down} \le \ell$ 
subtending $-\tfrac{\pi}{2} - \epsilon_\text{down} \ge \theta_\text{down} \ge \theta_\text{return}$, and flipping the sign of both the vertical coordinate $y$ and the shifted tension $\sigma - v^2$  of this portion of curve. 

It turns out that the above procedure for matching positions only works if the upper and lower portions of the curve are the same length, $\ell_\text{up} = \ell_\text{down} = \tfrac{\ell}{2} = \tfrac{1}{2}$, between the launch point and the vertical singularity.
This result is, surprisingly, independent of the launch and return angles, so that, for example, the lengths of any red and green curves in the family shown in \cite[figure 3]{Daerr19} are the same. 
This equal length condition was found analytically by Gregory \cite{Gregory49}.  
 We reproduce an equivalent demonstration in the following subsection. 

\subsection{Gregory's demonstration of equal upper and lower lengths}\label{gregory}

Borrowing the shorthand notation of \cite{Gregory49}, we define the quantity appearing in equations \eqref{eq:curvatureapp}-\eqref{eq:y} as $\tau \equiv \tan\left(\frac{\pi}{4}+\frac{\theta}{2}\right)$, and note that $\tau(-\tfrac{\pi}{2})=0$. 
Using (\ref{eq:s}-\ref{eq:y}) 
with a launch angle such that $-\frac{\pi}{2} \le \theta_1 = \theta_\text{launch} < \frac{\pi}{2}$ and $-\frac{\pi}{2}\leq\theta_2 \le\theta_\text{launch}$, we may obtain the shape of the upper portion of the loop ($s_\text{up}$, $x_\text{up}$, $y_\text{up}$) 
and find $C_\text{up}$ by inverting \eqref{eq:s} with a chosen $\theta_\text{launch}$ and $\ell_\text{up} = s(-\frac{\pi}{2})$.  Explicitly, we have 
\begin{align}
    C_\text{up} 
    = -\frac{1}{ 2 \ell_\text{up} } \left[ \frac{ \tau (\theta_\text{launch})^{D+1} }{D+1} + \frac{ \tau (\theta_\text{launch})^{D-1} }{D-1} \right] \, .
    \label{eq:Cup}
\end{align} 
We may obtain the shape of the lower portion of the loop ($s_\text{down}$, $x_\text{down}$, $y_\text{down}$) by first creating an upper loop such that $-\frac{\pi}{2} \le \theta_1 = \theta_\text{return}< \frac{\pi}{2}$ and $-\frac{\pi}{2}\leq\theta_2\leq\theta_\text{return}$,   
which will then be shifted and flipped as described in the previous subsection.   
We have 
\begin{align}
C_\text{down}  &= -\frac{1}{2 \ell_\text{down} } \left[ \dfrac{\tau \left( \theta_\text{return} \right)^{D+1}}{D+1} + \dfrac{\tau \left(\theta_\text{return} \right)^{D-1}}{D-1} \right] \, .
\end{align}
For a given $\theta_\text{launch}$ and $\ell_\text{up}$, $\theta_\text{return}$ and $C_\text{down}$ are found by patching the two curves at the vertical endpoints $\theta_2 = -\frac{\pi}{2}$ by setting $x_\text{up}\left(-\frac{\pi}{2}\right) = x_\text{down}\left(-\frac{\pi}{2}\right)$ and $y_\text{up}\left(-\frac{\pi}{2}\right) =- y_\text{down}\left(-\frac{\pi}{2}\right)$.  This process requires the following relations to hold: 
\begin{align}
\frac{C_\text{up}}{C_\text{down}} &=   \left[ \frac{ \tau(\theta_\text{launch}) }{ \tau(\theta_\text{return}) } \right]^D
\label{eq:Cdown} \, ,
\\
 \tau(\theta_\text{launch})\tau(\theta_\text{return}) &= \frac{D+1}{D-1} \, , 
\label{eq:thetareturn}
\end{align}
which, when inserted back into the the arc length equation \eqref{eq:s}, implies the equality of lengths
\begin{equation}
     \ell_\text{down}  
     = \ell_\text{up} \, . 
    \label{eq:LupLdown}
\end{equation}

\subsection{Numerical solution of heavy \emph{elastica}}\label{numelastica}

The heavy \emph{elastica} / stiff catenary equations were solved by projecting the linear momentum balance \eqref{eq:elasticaeom} onto $\bm{\hat{e}}_1$ and $\bm{\hat{e}}_2$, and solving alongside the angular momentum balance \eqref{eq:elasticaangular}. 
We solved for $\bm{f}(s)$, $\theta(s)$, $d_s\theta(s)$, and $\bm{x}(s)$ with a shooting method, using initial conditions $\bm{x}(0)$ and $\theta(0)$ and targets $\bm{x}(1)$ and $\theta(1)$. 
Other quantities may now be derived, including the tension, tangents, and normals.  The global force balance \eqref{eq:globalf} serves as a check on the validity of solutions; this balance and the target angle and position components are respectively satisfied to within $8 \times 10^{-11}$ and $9 \times 10^{-10}$ in the examples shown. 

\bibliographystyle{unsrt}  

\begin{thebibliography}{10}

\bibitem{IPT19}
11th {I}nternational {P}hysicists' {T}ournament 2019.
\newblock \url{https://2019.iptnet.info/problems/}.

\bibitem{YeanyYouTube}
B.~Yeany.
\newblock String shooter-string launcher- physics of toys.
\newblock \url{https://www.youtube.com/watch?v=rffAjZPmkuU}.

\bibitem{Gregory49}
C.~C.~L. Gregory.
\newblock Theory of a loop revolving in air, with observations on the
  skin-friction.
\newblock {\em The Quarterly Journal of Mechanics and Applied Mathematics},
  II(1):30--39, 1949.

\bibitem{SvetlitskiiMiroshnik73}
V.~A. Svetlitskii and R.~A. Miroshnik.
\newblock Critical velocities of the steady-state motion of an elastic fiber in
  a plane homogeneous flow.
\newblock {\em Soviet Applied Mechanics}, 9:542--545, 1973.

\bibitem{Miroshnik01}
R.~Miroshnik.
\newblock The phenomenon of steady-state string motion.
\newblock {\em Journal of Applied Mechanics}, 68:568--574, 2001.

\bibitem{ChakrabartiHanna16}
B.~Chakrabarti and J.~A. Hanna.
\newblock Catenaries in viscous fluid.
\newblock {\em Journal of Fluids and Structures}, 66:490--516, 2016.

\bibitem{StringLauncher}
B.~Yeany.
\newblock String launcher.
\newblock Yeany Educational Products, Palmyra, PA.

\bibitem{Aitken1878}
J.~Aitken.
\newblock An account of some experiments on rigidity produced by centrifugal
  force.
\newblock {\em Philosophical Magazine}, 5(29):81--105, 1878.

\bibitem{LariatChain}
N.~Tuck.
\newblock Lariat chain.
\newblock \url{https://www.normantuck.com/lariatchain}, 1987.

\bibitem{HealeyPapadopoulos90}
T.~J. Healey and J.~N. Papadopoulos.
\newblock Steady axial motions of strings.
\newblock {\em Journal of Applied Mechanics}, 57:785--787, 1990.

\bibitem{Routh55}
E.~J. Routh.
\newblock {\em The Advanced Part of a Treatise on the Dynamics of a System of
  Rigid Bodies}.
\newblock Dover, New York, 1955.

\bibitem{Airy1858}
G.~B. Airy.
\newblock On the mechanical conditions of the deposit of a submarine cable.
\newblock {\em Philosophical Magazine}, 16:1--18, 1858.

\bibitem{Thomson1857both}
W.~H. Thomson.
\newblock On machinery for laying submarine telegraph cables.
\newblock {\em The Engineer}, 4:185--187 and 280, 1857.

\bibitem{Gray59}
A.~Gray.
\newblock {\em A Treatise on Gyrostatics and Rotational Motion}.
\newblock Dover, New York, 1959.

\bibitem{PerkinsMote89}
N.~C. Perkins and C.~D. Mote, Jr.
\newblock Theoretical and experimental stability of two translating cable
  equilibria.
\newblock {\em Journal of Sound and Vibration}, 128:397--410, 1989.

\bibitem{HannaKing11}
J.~A. Hanna and H.~King.
\newblock An instability in a straightening chain.
\newblock [arXiv:1110.2360].

\bibitem{HannaSantangelo12}
J.~A. Hanna and C.~D. Santangelo.
\newblock Slack dynamics on an unfurling string.
\newblock {\em Physical Review Letters}, 109:134301, 2012.

\bibitem{Biggins14}
J.~S. Biggins.
\newblock Growth and shape of a chain fountain.
\newblock {\em Europhysics Letters}, 106:44001, 2014.

\bibitem{GregoryObit1}
C.~W. Allen.
\newblock Obituary: Mr. {C}. {C}. {L}. {G}regory.
\newblock {\em Nature}, 205:342, 1965.

\bibitem{GregoryObit2}
E.~M. Burbidge.
\newblock Obituary {N}otice: Christopher {C}live {L}angton {G}regory.
\newblock {\em Quarterly Journal of the Royal Astronomical Society}, 7:81--83,
  1965.

\bibitem{JuddLariat}
G.~Judd.
\newblock Effect of skin drag and rotation effects on a spinning loop of line.
\newblock \url{https://vimeo.com/72049798}.

\bibitem{Kurkin64}
V.~I. Kurkin.
\newblock The steady-state motion of a flexible yarn.
\newblock {\em Technology of the Textile Industry, U.S.S.R.}, 1964(6):41--47,
  1964.

\bibitem{SvetlitskiiGabryuk66}
V.~A. Svetlitskii and V.~I. Gabryuk.
\newblock The critical velocity of steadystate motion.
\newblock {\em Soviet Applied Mechanics}, 2:77--78, 1966.

\bibitem{Svetlitskii72inrussian}
V.~A. Svetlitskii, R.~A. Miroshnik, and V.~I. Kurkin.
\newblock Determining the forms of stationary thread motion at an arbitrary
  launch angle.
\newblock {\em Izvestiya vysshikh uchebnykh zavedeniy. Mashinostroyeniye},
  1972(3):37--42, 1972.
\newblock [in Russian].

\bibitem{Svetlitskii72}
V.~A. Svetlitskii, R.~A. Miroshnik, and V.~I. Kurkin.
\newblock Steady-state motion of a filament in media of different viscosities.
\newblock {\em Soviet Applied Mechanics}, 8:423--426, 1972.

\bibitem{Miroshnik72inrussian}
R.~A. Miroshnik.
\newblock Investigation of the steady motion of a ballistic antenna in a planar
  homogeneous flow.
\newblock {\em Izvestiya vysshikh uchebnykh zavedeniy. Mashinostroyeniye},
  1972(10):27--32, 1972.
\newblock [in Russian].

\bibitem{MiroshnikKurkin76inrussian}
R.~A. Miroshnik and V.~I. Kurkin.
\newblock On the planar stationary motion of a flexible thread.
\newblock {\em Izvestiya vysshikh uchebnykh zavedeniy. Mashinostroyeniye},
  1976(4):9--14, 1976.
\newblock [in Russian].

\bibitem{Judddolphin}
G.~Judd.
\newblock Personal communication, 2013.

\bibitem{BevanDeane19}
J.~J. Bevan and J.~H.~B. Deane.
\newblock A rigorous mathematical treatment of the shape of a dissipative rope
  fountain.
\newblock {\em Proceedings of the Royal Society A}, 475:20190612, 2019.

\bibitem{Abello20}
M.~Abello, J.~Courson, A.~Maury, and J.~Renaud.
\newblock {S}tring {S}hooter's overall shape in ambient air.
\newblock {\em Emergent Scientist}, 4:1--8, 2020.

\bibitem{Taberlet19}
N~Taberlet, J~Ferrand, and N~Plihon.
\newblock Propelled strings: Rising from friction.
\newblock {\em Physical Review Letters}, 123:144501, 2019.

\bibitem{Daerr19}
A~Daerr, J~Courson, M~Abello, W~Toutain, and B~Andreotti.
\newblock The charmed string: self-supporting loops through air drag.
\newblock {\em Journal of Fluid Mechanics}, 877:R2, 2019.

\bibitem{xkcd}
R.~Munroe.
\newblock Duty calls.
\newblock \url{https://xkcd.com/386/}.

\bibitem{MillerYouTube}
J.~Miller.
\newblock Self siphoning bead ball-chain - from a parking garage roof.
\newblock \url{https://www.youtube.com/watch?v=_ZhJFCSIzR8}.

\bibitem{OReilly17book}
O.~M. O'Reilly.
\newblock {\em Modeling Nonlinear Problems in the Mechanics of Strings and
  Rods}.
\newblock Springer, New York, 2017.

\bibitem{WatsonWang81}
L.~T. Watson and C.~Y. Wang.
\newblock Hanging an elastic ring.
\newblock {\em International Journal of Mechanical Sciences}, 23:161--167,
  1981.

\bibitem{Wang86}
C.~Y. Wang.
\newblock A critical review of the heavy elastica.
\newblock {\em International Journal of Mechanical Sciences}, 28:549--559,
  1986.

\bibitem{MaddocksDichmann94}
J.~H. Maddocks and D.~J. Dichmann.
\newblock Conservation laws in the dynamics of rods.
\newblock {\em Journal of Elasticity}, 34:83--96, 1994.

\bibitem{SinghHanna21}
H.~Singh and J.~A. Hanna.
\newblock Pseudomomentum: origins and consequences.
\newblock {\em Zeitschrift f{\"{u}}r angewandte Mathematik und Physik}, 72:122,
  2021.

\bibitem{Broer70}
L.~J.~F. Broer.
\newblock On the dynamics of strings.
\newblock {\em Journal of Engineering Mathematics}, 4:195--202, 1970.

\bibitem{Healey96}
T.~J. Healey.
\newblock Stability of axial motions of nonlinearly elastic loops.
\newblock {\em Zeitschrift f{\"{u}}r angewandte Mathematik und Physik},
  47:809--816, 1996.

\bibitem{Plunkett67}
R.~Plunkett.
\newblock Static bending stresses in catenaries and drill strings.
\newblock {\em Journal of Engineering For Industry}, 89:31--36, 1967.

\bibitem{StumpvanderHeijden00}
D.~M. Stump and G.~H.~M. van~der Heijden.
\newblock Matched asymptotic expansions for bent and twisted rods: applications
  for cable and pipeline laying.
\newblock {\em Journal of Engineering Mathematics}, 38:13--31, 2000.

\bibitem{Wolfe93}
P.~Wolfe.
\newblock Hanging cables with small bending stiffness.
\newblock {\em Nonlinear Analysis, Theory, Methods \& Applications},
  20(10):1193--1204, 1993.

\end{thebibliography}

\end{document}